\documentclass{PoS}
\usepackage[T1]{fontenc} 
\usepackage{textcomp}
\usepackage{slashed}
\usepackage{graphicx}%
\usepackage{subfigure}
\usepackage{mathptmx}
\usepackage{mathrsfs}
\usepackage{bm}
\usepackage{verbatim}
\newcommand{\beq}{\begin{equation}}
\newcommand{\eeq}{\end{equation}} 
\newcommand{\be}{\begin{eqnarray}}
\newcommand{\ee}{\end{eqnarray}}
\long\def\hidestart#1\hideend{}
\def\be{\begin{eqnarray}}
\def\ee{\end{eqnarray}}
\def\bea{\begin{eqnarray}}
\def\eea{\end{eqnarray}}
\def\ol{\overline}
\def\nn{\nonumber}

\def\lmatrix{\left(\begin{array}}
\def\rmatrix{\end{array}\right)}
\def\ee{\end{eqnarray}}
\def\tr{{\rm Tr\,}}

\title{Baryon spectrum in the composite sextet model}

\ShortTitle{Sextet baryons}

\author{Zoltan Fodor\\
        University of Wuppertal, Department of Physics, Wuppertal D-42097, Germany\\
        Juelich Supercomputing Center, Forschungszentrum Juelich, Juelich D-52425, Germany\\
        Eotvos University, Pazmany Peter setany 1, 1117 Budapest, Hungary\\
        \email{fodor@bodri.elte.hu}}

\author{Kieran Holland\\
        University of the Pacific, 3601 Pacific Ave, Stockton CA 95211, USA\\
        \email{kholland@pacific.edu}}

\author{Julius Kuti\\
        University of California, San Diego, 9500 Gilman Drive, La Jolla, CA 92093, USA\\
        \email{jkuti@ucsd.edu}}

\author{\speaker{Santanu Mondal}\\
        Eotvos University, Pazmany Peter setany 1, 1117 Budapest, Hungary\\
        MTA-ELTE Lendulet Lattice Gauge Theory Research Group, 1117 Budapest, Hungary\\
        \email{santanu@bodri.elte.hu}}

\author{Daniel Nogradi\\
        Eotvos University, Pazmany Peter setany 1, 1117 Budapest, Hungary\\
        MTA-ELTE Lendulet Lattice Gauge Theory Research Group, 1117 Budapest, Hungary\\
        \email{nogradi@bodri.elte.hu}}

\author{Chik Him Wong\\
        University of California, San Diego, 9500 Gilman Drive, La Jolla, CA 92093, USA\\
        University of Wuppertal, Department of Physics, Wuppertal D-42097, Germany\\
        \email{cwong@uni-wuppertal.de}}

\abstract{The strongly coupled near-conformal gauge theory with two fermion
    flavors in the two-index symmetric (sextet) representation of $SU(3)$
    is potentially a minimal realization of the composite Higgs
    mechanism. 
    We discuss the staggered fermion construction of 
    baryonic states, present our first numerical results and comment on
    implications for dark matter.}

\FullConference{The 32nd International Symposium on Lattice Field Theory,\\
		23-28 June, 2014\\
		Columbia University New York, NY}

\begin{document}

\section{Introduction}
\label{sec:intro}

$SU(3)$ gauge theory with two massless flavors of fermions in the 2-index symmetric
(sextet) representation may give rise to a minimal model of a composite Higgs particle
\cite{Dietrich:2005jn,Sannino:2004qp,Hong:2004td}.
The composite scalar, perhaps a Higgs impostor, could be a light and narrow state if
the underlying theory is near-conformal. 
There is accumulating non-perturbative evidence of this model being near-conformal
\cite{Fodor:2012ty,Fodor:2012ni,Fodor:2014pqa}, consistent with finite-temperature studies 
\cite{Kogut:2011ty,Kogut:2010cz} and also
with a small and non-zero $\beta$-function of the renormalized gauge coupling
\cite{DeGrand:2010na, Shamir:2008pb}. Furthermore, our recent results indeed show the presence of a light
composite scalar with $0^{++}$ quantum numbers in the spectrum of the theory \cite{Fodor:2012ni, Fodor:2014pqa}.
In all of these lattice studies, it should be noted, not all systematic effects are controlled
and hence should be interpreted with caution. 

In this model spontaneous symmetry breaking ${SU(2)_L
\times SU(2)_R\longrightarrow SU(2)_V}$ generates exactly three Goldstone bosons which are eaten to 
give masses to the ${ W^{\pm}}$ and $Z^0$ gauge bosons when the electroweak interaction is turned on.
There are no leftover massless or very light particles from the Goldstone spectrum which has implications for model
building regarding dark matter. In one class of composite Higgs models (some of) the remnant Goldstone bosons
become dark matter candidates \cite{Lewis:2011zb} which in our model is not possible. Nevertheless, a
stable baryon state is present in the spectrum which will be the topic of this contribution. 
The nature of
the baryonic state and its connection to dark matter require the coupling
of the sextet theory to the rest of the Standard Model to be specified.

The two flavors will be denoted by $\psi_{ab} = (u_{ab}, d_{ab})$ in analogy with QCD, they carry two $SU(3)$ gauge indices $a,b =
1,2,3$. The left-handed projections are a weak isospin doublet and the right-handed projections
are two weak isospin singlets,
\be
\psi_{ab}^L = \lmatrix{c} u_{ab}^L \\ d_{ab}^L \rmatrix \;, \qquad \psi_{ab}^R = \lmatrix{cc} u_{ab}^R , d_{ab}^R \rmatrix \;.
\ee
In order to not generate a $U(1)$ anomaly the cubic expression involving the flavor charges must vanish as usual. This
requirement is of course fullfilled for the $SU(2)$ generators acting on the left-handed doublet (Pauli matrices) and the non-trivial
constraint comes from the $U(1)$ hypercharge, $\tr Y = 0$. We choose $Y=2(Q-T_3)$ as the hypercharge generator, where
$Q$ is the electromagnetic charge and $T_3$ is the third component of weak isospin. Then an anomaly free setup requires
fractional charges for the left-handed doublet, $Q(u_L) = 1/2$ and $Q(d_L) = -1/2$. For the right-handed fermions, which
are electroweak singlets, we have hypercharge $Y=1$ for $u_R$ and $Y=-1$ for $d_R$, leading to a consistent charge assignment $Q(u_R)
= 1/2$ and $Q(d_R) = -1/2$. The anomaly condition from $\tr Y^3$ is also satisfied as can be checked directly.

The chiral symmetry group $SU(2)_L \times SU(2)_R$ breaks to the diagonal $SU(2)_V$ and hence $SU(2)_W \times
U(1)_Y$ breaks to $U(1)_{em}$. The surviving $U(1)_{em}$ symmetry ensures that baryon number is
conserved and also that the lightest baryon state is stable under the new gauge force and weak decay. 

There are two baryon states forming an isospin doublet of the type $(uud,udd)$ and they carry half-integer charge
with opposite sign. If these are to play any role as a dark matter particle it hence will be
of the fractionally charged massive particle (FCHAMP) type \cite{DeRujula:1989fe, Langacker:2011db}. 
As we argue in \cite{ourfuture}, the abundance of relic sextet baryons from the evolution history of a charge
symmetric Universe is expected to remain far below detectable levels. There are ways to modify the model such that more
contact can be made with direct detection experiments but these would compromise the minimality of the sextet model.
More importantly the viability of the model first and foremost hinges on the question whether an appropriate composite
Higgs particle resides in the spectrum or not. Nevertheless further speculation about the role the sextet baryons can
play as dark matter candidates can be found in \cite{ourfuture}.

\vspace{-0.3cm}

\section{Construction of the sextet nucleon operator}

\vspace{-0.2cm}

In the first two parts of this section we discuss the color, spin and flavor structure of the sextet nucleon
state in the continuum. 
We will see that a symmetric color contraction is needed
in order to construct a color singlet three fermion state when fermions are in the sextet representation. This makes the 
construction of the nucleon operator non-trivial in the lattice staggered basis, 
which we discuss in the third part of this section. 

\vspace{-0.1cm}

\subsection{Color structure}

\vspace{-0.1cm}

Three $SU(3)$ sextet fermions can give rise to a color singlet. The tensor product $6 \otimes 6 \otimes 6$ can be decomposed 
into irreducible representations of $SU(3)$ as \cite{maria}, 
\be
6 \otimes 6 \otimes 6= 1\oplus2\times 8\oplus 10\oplus \ol{10}\oplus 3\times 27\oplus 28\oplus 2\times 35
\ee
where irreps are denoted by their dimensions and $\ol{10}$ is the complex conjugate of 10.
The color singlet state corresponds to the unique singlet above. Fermions in the $6$-representation carry 2 indices,
$\psi_{ab}$, are symmetric, and transform as
\be
\psi_{aa'} \longrightarrow U_{ab}~U_{a'b'}~\psi_{bb'}
\ee 
and the singlet can be constructed explicitly as
\be
\epsilon_{abc}~\epsilon_{a'b'c'}~\psi_{aa'}~ \psi_{bb'}~ \psi_{cc'}\;.
\ee 
Let us introduce the index $A=0,\ldots5$ for the $6$ components of the symmetric $\psi_{ab}$, i.e. switch notation to
$\psi_{ab} = \Psi_A$. Then the above color singlet combination may be written as
\be
\epsilon_{abc}~\epsilon_{a'b'c'}~\psi_{aa'}~ \psi_{bb'}~ \psi_{cc'} = T_{ABC}~ \Psi_A~ \Psi_B~ \Psi_C
\label{colorsinglet}
\ee
with a totally symmetric 3-index tensor $T_{ABC}$. Note that in QCD the color contraction of the nucleon is
antisymmetric while for the sextet representation it is symmetric.

\vspace{-0.1cm}

\subsection{Spin flavor structure}

\vspace{-0.1cm}

As we have seen the color contraction is symmetric for the sextet representation and hence the overall antisymmetry
of the baryon wave function with respect to interchanging any two fermions in it must come from the spin-flavor structure.

It is useful to consider the non-relativistic notation and suppress color indices. The two flavors will be labelled as
$u$ and $d$ like in QCD and the non-relativistic spin will be either $\uparrow$ or $\downarrow$. The state we are after
may be obtained from $| \uparrow u, \uparrow d, \downarrow u \rangle$ by making it totally antisymmetric,
\be
\label{psi}
| \psi  \rangle = 
| \uparrow u, \uparrow d, \downarrow u \rangle + 
| \downarrow u, \uparrow u, \uparrow d \rangle + 
| \uparrow d, \downarrow u, \uparrow u \rangle - \nn \\
| \downarrow u, \uparrow d, \uparrow u \rangle - 
| \uparrow d, \uparrow u, \downarrow u \rangle - 
| \uparrow u, \downarrow u, \uparrow d \rangle\;.
\ee

\vspace{-0.1cm}

\subsection{From continuum Dirac to lattice Staggered basis}

\vspace{-0.1cm}

The lattice operators that create the state (\ref{psi})
belong to a suitable multiplet 
of taste $SU(4)$. Our staggered operator construction follows \cite{gs, KlubergStern:1983dg}. 

We motivate our staggered operator construction from the correct operator in Dirac basis.
For simplicity we want to have operators as local as possible, thus in Dirac basis, our sextet nucleon operator takes the form,
\be
N^{\alpha i}(2y)~=T_{ABC}~ u_A^{\alpha i}(2y) ~[u_B^{\beta j}(2y) ~(C \gamma_5 )_{\beta \gamma} 
~(C^* \gamma_5^*)_{ij}~d_C^{\gamma j}(2y)]
\ee
where Greek letters and lower case Latin letters
denote spin and taste indices, respectively. 
$C$ is the charge conjugation matrix satisfying
\be
C\gamma_\mu C^{-1}=-\gamma_\mu^{{\rm T}},\nn\\
-C=C^{{\rm T}}=C^\dagger=C^{-1}.
\ee
The coordinate $y$ labels elementary staggered hypercubes.
Staggered fields are defined as
\be
u^{\alpha i}(2y)=\frac{1}{8}\sum_\eta \Gamma^{\alpha i}_\eta~ \chi_u(2y+\eta)\;, \nn
\ee
where $\Gamma(\eta)$ is an element of the Euclidean Clifford algebra labeled by the four-vector 
$\eta$ whose elements are defined mod 2 as usual.
More precisely 
$\Gamma(\eta)=\gamma_1^{\eta_1}\gamma_2^{\eta_2}\gamma_3^{\eta_3}\gamma_4^{\eta_4}$ where 
$\eta \equiv (\eta_1, \eta_2, \eta_3, \eta_4)$.
Writing in terms of the staggered fields,
\be
N^{\alpha i}(2y)~=-T_{ABC}~\frac{1}{8^3}\sum_{\eta'} \Gamma^{\alpha i}_{\eta'}~ \chi_u^A(2y+\eta')
~\sum_{\eta }  S(\eta)\chi_u^B(2y+\eta) \chi_d^C(2y+\eta)\;,
\ee
where $S(\eta)$ is a sign factor.
To obtain a single time slice operator an extra term has to be added to or subtracted from the diquark operator to cancel
the spread over two time slices of the unit hypercube. This is similar to what is done to construct 
the single time slice meson operators in QCD. This extra term corresponds to the parity partner of the nucleon.
The single time slice nucleon operator reads,
\be
N^{\alpha i}(2y)~=-T_{ABC}~\frac{1}{8^3}\sum_{{\vec \eta'}} \Gamma^{\alpha i}_{{\vec \eta'}}~ \chi_u^A(2y+{\vec \eta'})
~\sum_{{\vec \eta} }  S({\vec \eta})\chi_u^B(2y+{\vec \eta}) \chi_d^C(2y+{\vec \eta}).
\ee
Now the operator is a sum of $8\times 8=64$ terms over the elementary cube in a given time slice. 
The local terms vanish individually after the symmetric 
color contraction. The non-vanishing terms are those where a diquark resides on a corner of the cube at a fixed 
time-slice and the third fermion resides
on any of the other corners. The nucleon operator is thus the sum of all such $56$ terms with appropriate sign factors. 
In order to find the mass of the lowest lying state any one of these $56$ terms can in principle be used.
We use the operators listed in Table \ref{operators}.

\begin{table}
\begin{center}
\begin{tabular}{|l|l|l|} 
\hline
Label  &  Operators (set a) & Operators (set b)  \\
\hline
 ${\rm IV_{{\rm xy}}}$   & $\chi_u(1,1,0,0)~\chi_u(0,0,0,0)~\chi_d(0,0,0,0)$&$\chi_u(0,0,0,0)~\chi_u(1,1,0,0)~\chi_d(1,1,0,0)$\\
 \hline
 ${\rm IV_{{\rm yz}}}$   & $\chi_u(0,1,1,0)~\chi_u(0,0,0,0)~\chi_d(0,0,0,0)$&$\chi_u(0,0,0,0)~\chi_u(0,1,1,0)~\chi_d(0,1,1,0)$\\
 \hline
 ${\rm IV_{{\rm zx}}}$   & $\chi_u(1,0,1,0)~\chi_u(0,0,0,0)~\chi_d(0,0,0,0)$&$\chi_u(0,0,0,0)~\chi_u(1,0,1,0)~\chi_d(1,0,1,0)$\\
 \hline
 VIII   & $\chi_u(1,1,1,0)~\chi_u(0,0,0,0)~\chi_d(0,0,0,0)$&$\chi_u(0,0,0,0)~\chi_u(1,1,1,0)~\chi_d(1,1,1,0)$\\
 \hline
\end{tabular}
\end{center}
\caption{Staggered lattice operators used.}
\label{operators}
\end{table}  

\vspace{-0.3cm}

\section{Lattice simulations}

\vspace{-0.2cm}

The rooted staggered fermion action with 2-steps of stout-smearing 
\cite{mp} 
and tree-level Symanzik-improved gauge
action have been used to simulate two sextet flavors on the lattice. 
Simulations have been performed at one value of the bare coupling,
$\beta = 6 / g^2 = 3.2$. 

Autocorrelations are monitored by the time histories of effective masses and correlators. 
For the estimate of the statistical errors of hadron masses we used correlated fitting with double 
jackknife procedure on the covariance matrices \cite{delDebbio}.

\subsection{Nucleon operator comparison}

\begin{figure}
\begin{center}
\scalebox{0.55}{\includegraphics[angle=0,width=0.85\textwidth]{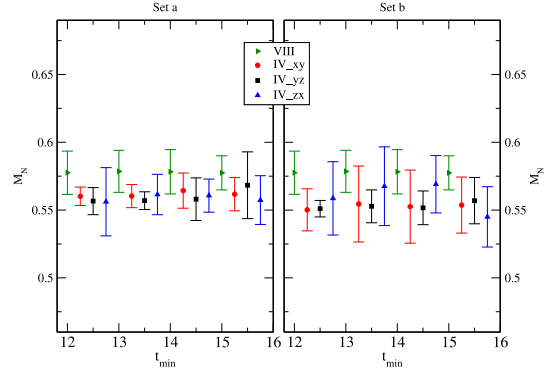}}
\vskip -.3 cm
\caption{{\footnotesize Comparison of $M_N$ from different operators varying $t_{min}$ 
with fixed $t_{max}=20$. The calculation is performed on lattices with $\beta=3.20$,
$V = 32^3 \times 64$ and $m=0.007$ using $200$ configurations. The $t_{min}$ 
values for the operators $IV_{xy}$, $IV_{yz}$ and $IV_{zx}$ are shifted by $0.25$,
$0.5$ and $0.75$ respectively for clarity.}}\label{signal_comparison}
\end{center}
\end{figure}

We investigate the signal qualities for the operators listed in Table \ref{operators} on
$200-300$ configurations (separated by 5 trajectories each) with volume $V=32^3 \times 64$ and fermion mass
$m=0.007$. For each operator the nucleon mass, $M_N$, is fitted from time
separation $t_{min}$ to $t_{max}$. Figure \ref{signal_comparison} compares the
corresponding fits for various values of $t_{min}$ at $t_{max}=20$. It is observed that,
for all operators, the fits are stable against the choices of fit ranges. Moreover, all
operators are consistent with one another within errors. Their noise-to-signal
ratios are similarly small and around $\sim 5\%$. There is no operator significantly
less noisy than the others, therefore the quality of the resulting spectroscopy would be
independent of the choice of operators. In the following analysis we use the operator
$IV_{xy}$ in set $a$.

\subsection{Preliminary results}

\begin{figure}
\begin{center}
\scalebox{0.55}{\includegraphics[angle=0,width=1.0\textwidth]{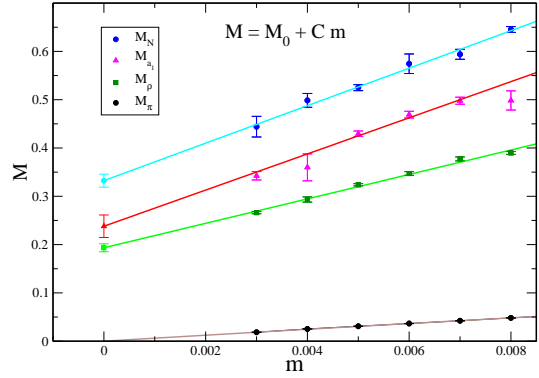}}
\vspace{-0.1cm}
\caption{
{\footnotesize Chiral extrapolation of $M_N$ (blue) in comparison with
$M_\pi$, $M_{a_1}$ and $M_\rho$. 
The calculation is performed on lattices with $\beta=3.20$, $V=48^3 \times 96$ 
for $m=0.003$ (except $M_N$ on $V=32^3 \times 64$) and $V=32^3 \times 64$ for 
the rest, using $200-300$ configurations.}
}
\label{chiral}
\end{center}
\end{figure}

In this section we present our preliminary results on the nucleon spectroscopy. The nucleon correlator of operator
$IV_{xy}$ in set $a$ is measured on lattices with volume $V = 32^3 \times 64$ and fermion mass $m=0.003$ to $m=0.008$, 
each with $200-300$ configurations.
Figure \ref{chiral} shows the chiral extrapolation of $M_N$ 
compared with the masses of pion, $a_1$ and $\rho$ mesons, denoted by
$M_\pi$, $M_{a_1}$ and $M_\rho$, respectively \cite{Fodor:2012ty, ourfuture}. 
It is observed that $M_N$ is heavier than the low-lying mesons in the chiral limit as expected.  

\begin{figure}
\begin{center}
\scalebox{0.45}{\includegraphics[angle=0,width=1.0\textwidth]{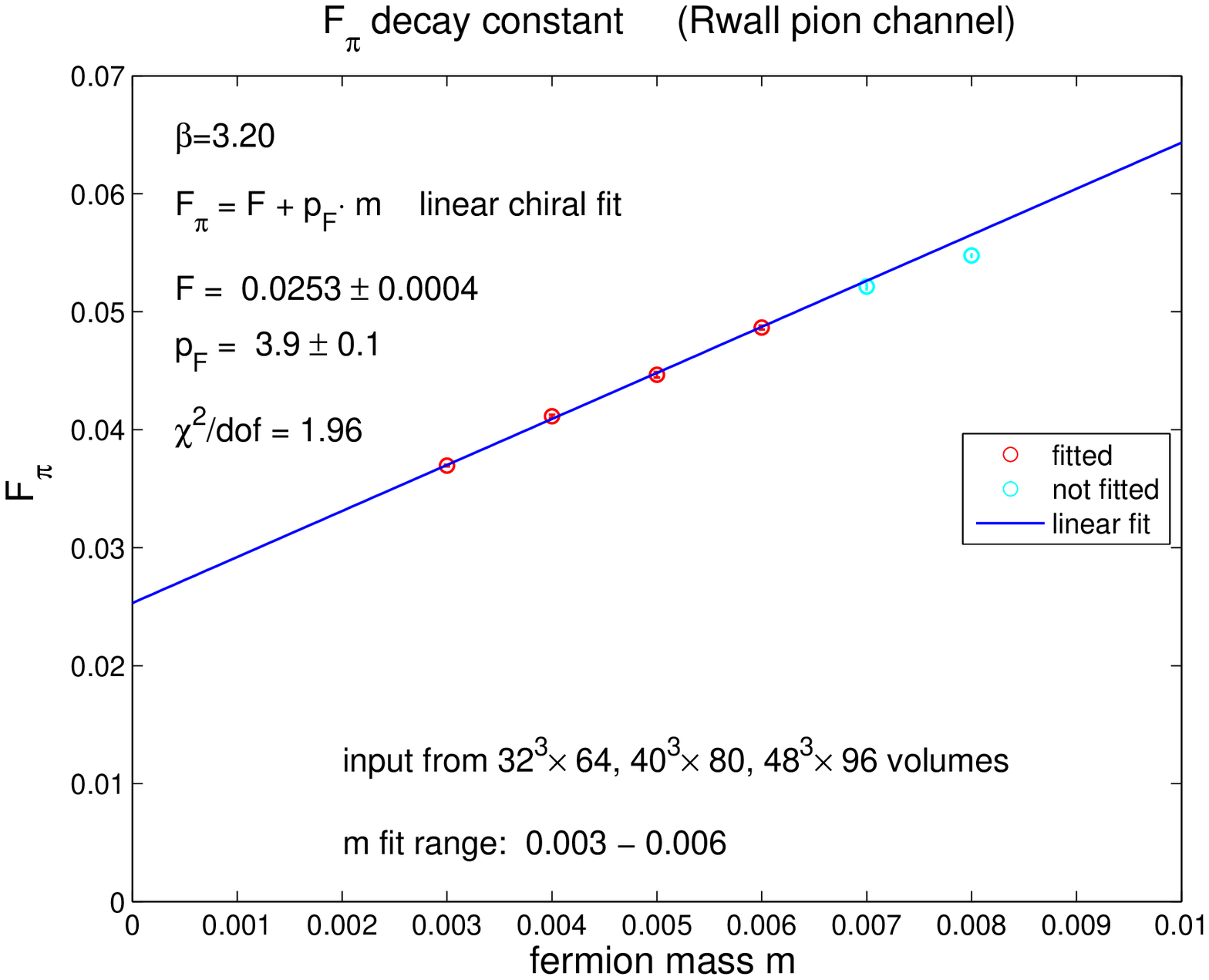}}
~ ~ ~
\scalebox{0.15}{\includegraphics[angle=0,width=1.0\textwidth]{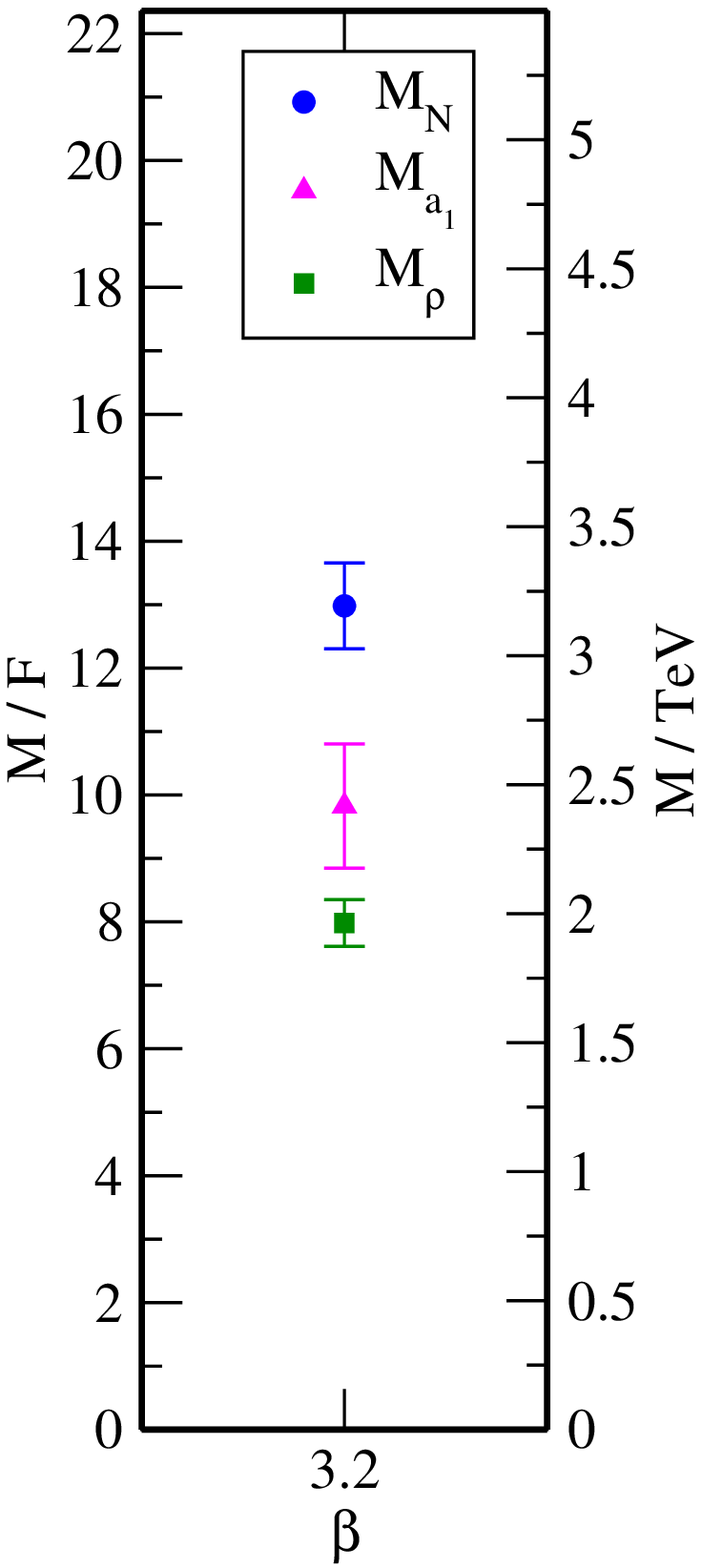}}
\vspace{-0.3cm}
\caption{{\footnotesize Preliminary chiral extrapolation of $F_\pi$. The calculation is performed on lattices with
$\beta=3.20$, $V=48^3 \times 96$ for $m=0.003$ and V$=32^3 \times 64$ for the rest, 
using $200-300$ configurations, see \cite{ourfuture} for more details (left). Hadron spectroscopy at $\beta=3.20$ in physical units (right).}}\label{physical}
\end{center}
\end{figure}

The scale is set by the chiral limit of $F_\pi$, denoted by $F$.  The left plot of figure \ref{physical} shows a
preliminary chiral
extrapolation of $F$, measured on lattices on $48^3 \times 96$ for $m=0.003$ and $32^3 \times 64$ for the
rest. As a preliminary result we obtain $F=0.0253(4)$; see \cite{ourfuture} for more details. 
Imposing $F=246\; GeV$, the nucleon is found to be approximately $3\; TeV$. It is compared with other meson states 
on the right panel of figure \ref{physical}.

\vspace{-0.3cm}

\section{Summary and outlook}

\vspace{-0.2cm}

We have determined the fermion mass dependence of a nucleon-like state in $SU(3)$ gauge theory coupled to two flavors
of massless fermions in the sextet representation. 
In our pilot study the chiral limit of the sextet nucleon mass is around
$3\; TeV$ at the particular lattice spacing we analysed. 

Our motivation for studying this particular nucleon-like
state is twofold. First, it has intriguing connections with our understanding of dark matter as discussed in
\cite{ourfuture}.
As a first step we studied
the mass of this new state on the $3\;TeV$ scale which is necessary for all dark matter considerations.
Our second motivation was the
expectation that regardless of what its physical interpretation is, the mass of the baryon state, or more precisely the
ratio of baryon states at different bare couplings compared with the ratio of other dimensionful quantities
can give indication how close or far we are from the continuum limit. So far we have analyzed $\beta = 3.20$ only
but this aspect will be
tested when we analyse our $\beta = 3.25$ lattices and compare the scale change from $\beta = 3.20$ coming from the ratio
of the baryon mass and other dimensionful quantities.


\vspace{-0.3cm}

\section{Acknowledgement}

\vspace{-0.2cm}


We acknowledge support by the DOE under grant DE-SC0009919,
by the NSF under grants 0970137 and 1318220, by the DOE ALCC award for the BG/Q Mira platform
of Argonne National Laboratory, by OTKA under the grant OTKA-NF-104034, and by the Deutsche
Forschungsgemeinschaft grant SFB-TR 55. Computational resources were provided by USQCD at Fermilab, 
by the NSF XSEDE program, by the University of Wuppertal, by Juelich Supercomputing Center on Juqueen
and by the Institute for Theoretical Physics,
We are grateful to Szabolcs Borsanyi for his code development for the BG/Q platform. We are also 
grateful to Sandor Katz and Kalman Szabo for their code developent for the CUDA platform \cite{Egri:2006zm}.



\vspace{-0.2cm}

\begin{thebibliography}{99}

\vspace{-0.2cm}

{\footnotesize

\bibitem{Dietrich:2005jn} 
D.~D.~Dietrich, F.~Sannino and K.~Tuominen,
Phys.\ Rev.\ D {\bf 72}, 055001 (2005)
[hep-ph/0505059].
\vspace{-0.1cm}

\bibitem{Sannino:2004qp} 
F.~Sannino and K.~Tuominen,
Phys.\ Rev.\ D {\bf 71}, 051901 (2005)
[hep-ph/0405209].
\vspace{-0.1cm}

\bibitem{Hong:2004td} 
D.~K.~Hong, S.~D.~H.~Hsu and F.~Sannino,
Phys.\ Lett.\ B {\bf 597}, 89 (2004)
[hep-ph/0406200].
\vspace{-0.1cm}

\bibitem{Fodor:2012ty} 
Z.~Fodor, K.~Holland, J.~Kuti, D.~Nogradi, C.~Schroeder and C.~H.~Wong,
Phys.\ Lett.\ B {\bf 718}, 657 (2012)
[arXiv:1209.0391 [hep-lat]].
\vspace{-0.1cm}

\bibitem{Fodor:2012ni} 
Z.~Fodor, K.~Holland, J.~Kuti, D.~Nogradi, C.~Schroeder and C.~H.~Wong,
PoS LATTICE {\bf 2012}, 024 (2012)
[arXiv:1211.6164 [hep-lat]].
\vspace{-0.1cm}

\bibitem{Fodor:2014pqa} 
Z.~Fodor, K.~Holland, J.~Kuti, D.~Nogradi and C.~H.~Wong,
PoS LATTICE {\bf 2013}, 062 (2014)
[arXiv:1401.2176 [hep-lat]].
\vspace{-0.1cm}


\bibitem{Kogut:2011ty} 
J.~B.~Kogut and D.~K.~Sinclair,
Phys.\ Rev.\ D {\bf 84}, 074504 (2011)
[arXiv:1105.3749 [hep-lat]].
\vspace{-0.1cm}

\bibitem{Kogut:2010cz} 
J.~B.~Kogut and D.~K.~Sinclair,
Phys.\ Rev.\ D {\bf 81}, 114507 (2010)
[arXiv:1002.2988 [hep-lat]].
\vspace{-0.1cm}

\bibitem{DeGrand:2010na} 
T.~DeGrand, Y.~Shamir and B.~Svetitsky,
Phys.\ Rev.\ D {\bf 82}, 054503 (2010)
[arXiv:1006.0707 [hep-lat]].
\vspace{-0.1cm}

\bibitem{Shamir:2008pb} 
Y.~Shamir, B.~Svetitsky and T.~DeGrand,
Phys.\ Rev.\ D {\bf 78}, 031502 (2008)
[arXiv:0803.1707 [hep-lat]].
\vspace{-0.1cm}

\bibitem{Lewis:2011zb} 
R.~Lewis, C.~Pica and F.~Sannino,
Phys.\ Rev.\ D {\bf 85}, 014504 (2012)
[arXiv:1109.3513 [hep-ph]].
\vspace{-0.1cm}

\bibitem{DeRujula:1989fe} 
A.~De Rujula, S.~L.~Glashow and U.~Sarid,
Nucl.\ Phys.\ B {\bf 333}, 173 (1990).

\bibitem{Langacker:2011db} 
P.~Langacker and G.~Steigman,
Phys.\ Rev.\ D {\bf 84}, 065040 (2011)
[arXiv:1107.3131 [hep-ph]].



\bibitem{ourfuture} 
Z.~Fodor, K.~Holland, J.~Kuti, S.~Mondal, D.~Nogradi and C.~H.~Wong,
PoS LATTICE {\bf 2014}, 244 (2014)
\vspace{-0.1cm}
 

\bibitem{appelquist} 
T.~Appelquist {\it et al.}  [Lattice Strong Dynamics (LSD) Collaboration],
Phys.\ Rev.\ D {\bf 88}, no. 1, 014502 (2013).
\vspace{-0.1cm}

\bibitem{maria}
Maria~S~M~Wessl\'{e}n,
Journal\ of\ Physics:\ Conference\ Series\ {\bf175}, (2009) 012015
\vspace{-0.1cm}

\bibitem{gs}
M.~F.~L.~Golterman and J.~Smit,
Nucl.\ Phys.\ B {\bf 255}, 328 (1985).
\vspace{-0.1cm}

\bibitem{KlubergStern:1983dg} 
H.~Kluberg-Stern, A.~Morel, O.~Napoly and B.~Petersson,
Nucl.\ Phys.\ B {\bf 220}, 447 (1983).
\vspace{-0.1cm}

\bibitem{mp} 
C.~Morningstar and M.~J.~Peardon,
Phys.\ Rev.\ D {\bf 69}, 054501 (2004).
\vspace{-0.1cm}

\bibitem{delDebbio} 
L.~Del Debbio, L.~Giusti, M.~Luscher, R.~Petronzio and N.~Tantalo,
JHEP {\bf 0702}, 082 (2007).

\bibitem{Egri:2006zm} 
G.~I.~Egri, Z.~Fodor, C.~Hoelbling, S.~D.~Katz, D.~Nogradi and K.~K.~Szabo,
Comput.\ Phys.\ Commun.\  {\bf 177}, 631 (2007)
[hep-lat/0611022].

}



\end{thebibliography}
\end{document}